\documentclass[twocolumn,floats,showpacs,aps,superscriptaddress,floatfix]{revtex4}
\usepackage{times} 
\usepackage{graphicx}
\usepackage{textcomp}
\begin{document}
\bibliographystyle{$HOME/Literature/Bibdir/revtex}
%
\title{Alternative mechanisms of structuring biomembranes: Self-assembly vs.
self-organization}
\author{Karin John}
\email{john@mpipks-dresden.mpg.de}
\affiliation{Max-Planck-Institut
f\"ur Physik komplexer Systeme, N{\"o}thnitzer Str.\ 38, D-01187 Dresden,
Germany}
\author{Markus B{\"a}r}
\email{markus.baer@ptb.de}
\affiliation{Physikalisch-Technische Bundesanstalt, Abbestr.\ 2--12, D-10587 Berlin, Germany}
\begin{abstract}
We study two mechanisms for the formation of protein patterns near
membranes of living cells by mathematical modelling.  Self-assembly
of protein domains by electrostatic lipid-protein interactions is
contrasted with self-organization due to a nonequilibrium biochemical
reaction cycle of proteins near the membrane. While both processes lead
eventually to quite similar patterns, their evolution occurs on very
different length and time scales.  Self-assembly produces periodic
protein patterns on  a spatial scale below  0.1\,\textmu{}m in a few seconds
followed by extremely slow coarsening, whereas self-organization results
in a pattern wavelength comparable to the typical cell size of 100\,\textmu{}m
within a few minutes suggesting different biological functions for the
two processes.
\end{abstract}
%
%
%
%
%
\maketitle

%
%
Living cells display internal structures on various
lenght scales, that are regulated dynamically.
Examples include the organisation of nerve cells into axon, body and
dendrites \cite{Alberts} or the occurrence of short-lived signaling patches
in the membrane of chemotaxing amoebae \cite{Manahan:2004}. 
The origin of many of those structures is a
nonuniform distribution of biochemical molecules, that can be achieved
by a spontaneous symmetry breaking through local fluctuations.
The diversity of time and length scales in cellular structures strongly
suggest that a variety of mechanisms participate in the structuring
process.
From a physics perspective, one can distinguish at least two different
classes of processes: self-assembly and self-organization
\cite{Misteli:2001}.

{\it Self-assembly} implies spatial structuring as a result of
minimization of the free energy in a closed system.
Hence, a self-assembled structure corresponds to a thermodynamic
equilibrium.  
A prominent example for self-assembly  in single cells is phase
separation of lipids and proteins due to macromolecular interactions.
Phase separation occurs if the interaction energies dominate the entropy
contribution.
Model membranes show a phase separation due to lipid-lipid interactions
\cite{Veatch:2002,
McConnell:2003} or lipid-protein interactions \cite{Glaser:1996} (for a
critical discussion see also \cite{Murray:1999}).
Theoretical analysis of phase separation in biomembranes has been
mainly
restricted to free energy considerations \cite{May:2002, Anderson:2000},
which can predict the final equilibrium state, but do not capture the
transient dynamics.

{\it Self-organization}, in contrast, requires a
situation far away from thermodynamic equilibrium and is possible only in
open systems with an external energy source.  
One prominent example is pattern
formation in reaction-diffusion systems, which has first been proposed by
Turing \cite{Turing:1952} and later on become influential in development biology 
 \cite{Gierer:1972}.
The key ingredients are nonlinear self-enhancing 
reactions, a supply of chemical energy and competing diffusion of the involved molecules.
Experimental patterns in single cells that are successfully modelled
by reaction-diffusion equations include calcium waves \cite{Falcke:2004}
and protein distributions in E. coli \cite{Howard:2005}.

In the vicinity of membranes  molecular interactions and
reaction-diffusion processes occur simultaneously. 
In this Letter we discuss and compare two dynamical models for the
alternative mechanisms of pattern formation near the cell membrane:
model~I describes self-assembly of protein domains due to lipid-protein
interactions and model~II describes an active reaction-diffusion
mechanism resulting in self-organization of proteins.

The two models are based on the properties of
the GMC proteins reviewed in \cite{Blackshear:1993,McLaughlin:2002}, which play a critical role in the development of the
neural system \cite{Frey:2000} and in
the regulation of cortical actin-based structures and cell motility
\cite{Laux:2000}.
A key result is that both models display similar qualitative
dynamical behavior but act on largely different time and length scales.
Model~I leads to the formation of stationary domains on a submicrometer
scale
within seconds. A slow coarsening process follows the expected power law.
Phase separation arises from a relaxational process into thermodynamic
equilibrium, where the equilibrium state is characterized by vanishing
fluxes.
Model~II leads to the formation of large stationary
structures on the scale of a eukaryotic cell within 10\,min. The steady
state
of this active phase separation (= phase separation from
self-organization) is characterized by nonzero fluxes.
%

\begin{figure}[hbt]
\includegraphics[width=0.7\hsize]{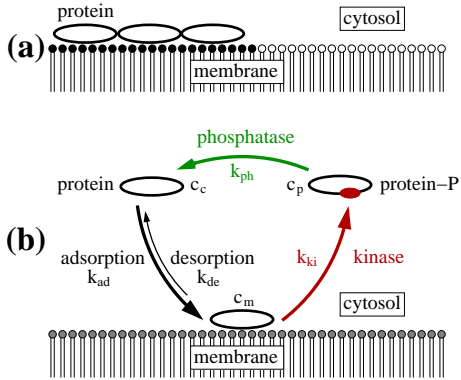}
\caption{Schematic depiction of models~I (a) and II (b). (a) The membrane consists of a mixture of type 1 (black) and 2 (white) lipids with
adsorbed proteins. (b) Proteins undergo a cycle of phosphorylation and
dephosphoylation as explained in the text. \label{mod}}. 
\end{figure}

%
%
%
%
{\bf \it Model~I} is based on attractive interactions of GMC proteins with
acidic lipids in membranes \cite{AWWH00}. In this mechanism (Fig.\,\ref{mod}\,(a)) proteins
(area fraction $c_m$) are associated with a lipid membrane, consisting of
type~1 (area fraction $c_l$) and type~2 (area fraction $1-c_l$) lipids of
identical size.
The size
ratio between proteins and lipids is $N$.
Proteins interact attractively with type~1 lipids. Proteins and lipids are
allowed to diffuse in
the plane of the membrane. 
The number of
bound proteins and the average membrane composition are conserved quantities.   
We write the free energy of the protein covered membrane similar to
\cite{May:2002} as
\begin{equation}
F   =
k_BTN_a\int\,dA\Bigg[f_l+\frac{\chi}{2}\left(\nabla{}c_l\right)^2\Bigg]\label{f1}
\end{equation}
where $dA$ denotes integration over the membrane surface, $N_a$ denotes the
number of lipids per unit area and
\begin{eqnarray}
f_l & = &c_l\ln{}c_l+(1-c_l)\ln{}(1-c_l)+\nonumber\\
   &   &\frac{c_m}{N}\ln{}c_m+\frac{1-c_m}{N}\ln{}(1-c_m)-uc_lc_m \label{f2}\,.
\end{eqnarray}
The local part of the free energy (\ref{f2}) consists of the entropic contributions of
the lipid (first and second term) and the protein phase (third
and fourth term) and the interaction energy between lipids
and proteins (last term) \cite{par:u}. For simplicity we assume that electrostatic repulsion and non-electrostatic attraction in the lipid and protein phase cancel locally. This is
somewhat arbitrary, but does not change the nature of the results as long as demixing is governed by lipid-protein interactions. 
The non-local part of the free energy (\ref{f1}) is governed by the interfacial
energy in the lipid phase, which is a small quantity.
For the parameter $\chi$ we set $\chi\approx{}\frac{1}{2}u_ll^2$ from
Cahn-Hilliard Theory \cite{Cahn:1958} where $u_l$ and $l$ are the
typical interaction energy (on the order of several $k_{\rm B}T$) and length (on the order of
a lipid headgroup size, i.e.\,$l\approx{}1-2$\,nm), respectively.  
Evolution equations for $c_l$ and $c_m$ are obtained using linear
nonequilibrium thermodynamics and the mass balance equations to give
\begin{eqnarray}
\begin{array}{lclclcl}
\partial_tc_l & = & -\nabla\cdot\vec{j}_l  & \quad \mbox{with}\quad & \vec{j}_l & = & -M_l\nabla\frac{\delta{}F}{\delta{}c_l}
\label{fluxl}\\
\partial_tc_m & = & -\nabla\cdot\vec{j}_m  & \quad \mbox{with}\quad &
\vec{j}_m & = & -M_m\nabla\frac{\delta{}F}{\delta{}c_m} \label{fluxm}\,.
\end{array}
\end{eqnarray}
More specifically, the fluxes $\vec{j}_l$ and $\vec{j}_m$ are given by
\begin{eqnarray}
\vec{j}_l & = & -D_l\left[\nabla c_l - c_l(1-c_l)\left(u\nabla c_m+\chi \nabla^3c_l\right)\right]\\
\vec{j}_m & = & -D_m\left[\nabla c_m - Nc_m(1-c_m)u\nabla c_l\right]\,,
\end{eqnarray}
where $D_l=M_lk_BTN_a$ and $D_m=M_mk_BTN_a$.

%
%
%
{\bf \it Model~II} is based on a biochemical cycle of GMC phosphorylation
and dephosphorylation, called myristoyl-electrostatic (ME) switch \cite{Thelen:1991}.
In contrast to model~I, we will assume that all different lipid species are
distributed uniformely in the membrane.
In this phase separation model (Fig.\,\ref{mod}\,(b)) a protein can associate
reversibly with the membrane following a mass action law.
On the membrane proteins are irreversibly phosphorylated by a protein
kinase, which disrupts membrane binding immediately and translocates the protein into the
cytosol where it is dephosphorylated by a phosphatase.
We can cast these
processes into a three-variable reaction-diffusion model of the following form
\begin{eqnarray}
\partial_tc_m & = &
k_{ad}(1-c_m)c_c-k_{de}c_m-\nonumber\\
& &k_{ki}(1-c_m)\frac{c_m}{k_m+c_m}+D_m\nabla^2c_m \label{sys1}\\
\partial_tc_c & = &k_{de}c_m-k_{ad}(1-c_m)c_c+k_{ph}c_p+D_c\nabla^2c_c\\
\partial_tc_p & = & k_{ki}(1-c_m)\frac{c_m}{k_m+c_m}-k_{ph}c_p+D_c\nabla^2c_p \label{sys3}\,.
\end{eqnarray}
$c_m$, $c_c$ and $c_p$ denote the concentrations of membrane bound, cytosolic unphosphorylated
and cytosolic phosphorylated proteins, respectively. $k_{ad}$ and $k_{de}$
denote the rate constants of membrane association and dissociation of the
unphosphorylated protein and $k_{ki}$ and $k_{ph}$ denote the enzymatic
activities of the kinase and phosphatase.
For the kinase activity we have used a Michaelis-Menten type kinetics \cite{Verghese:1994},
whereas we have neglected this property for the phosphatase, assuming that the
concentration of phosphorylated proteins is well below the respective
Michaelis-Menten constant \cite{Seki:1995}.
Based on the properties of protein kinase C we assume, that the kinase needs lipids for full activation \cite{Cockcroft:2000,Bell:1991} and that membrane bound proteins decrease the available membrane space and thus the
kinase activity \cite{Glaser:1996}.
Eqs.\,(\ref{sys1})--(\ref{sys3}) constitute a non-equilibrium system, since
the kinase activity is sustained and consumes ATP which is produced by the metabolism.
The total protein concentration $c_t=\frac{1}{V}\int{}dV\,(c_m+c_p+c_c)$ is
conserved.

\begin{figure}[hbt]
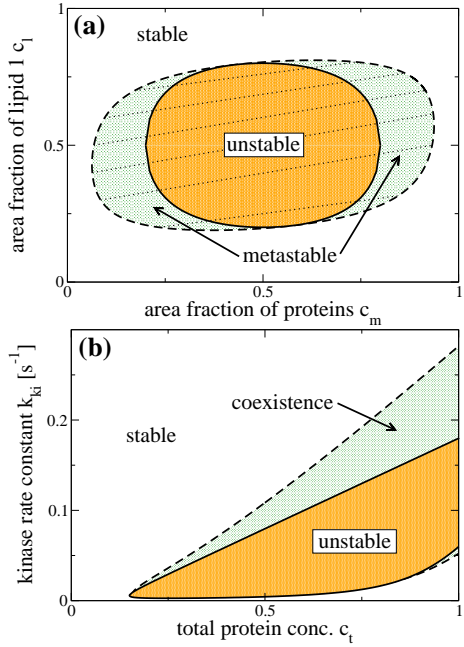

\includegraphics[width=0.7\hsize]{fig2a.eps}
\includegraphics[width=0.7\hsize]{fig2b.eps}
\caption{(a) Phase diagrams with tie-lines for the phase separation
in model~I and (b) phase diagram for the active phase separation
in model~II. Parameters in (a) are $N=25$ and $u=1.0$ and in (b)
are $N=25$, $D_m=0.04\,\mu$m$^2$\,s$^{-1}$ \cite{Swaminathan:1997,Seksek:1997}, $D_c=20\,\mu$m$^2$\,s$^{-1}$,
$k_{ad}=1$\,s$^{-1}$ \cite{Arbuzova:1997},
$k_{de}=0.005$\,s$^{-1}$ \cite{Arbuzova:1997}, $k_m=0.01$
\cite{Verghese:1994}, $k_{ph}=0.2$\,s$^{-1}$ \cite{Thelen:1991}.\label{phase}}. 
\end{figure}

%
%
The parameters for both models have been taken from experiments.
First we consider the linear stability of the uniform steady states
(Fig.\,\ref{phase}). In both models we can identify a region of linear
instability, characterized by real positive eigenvalues.
Using the Maxwell construction and tie-lines we calculated for model~I the
equilibrium state for a given uniform state and identified a metastable
region, where a stable uniform state coexists with a stable demixed state.  
The demixed state is
characterized by a phase with a high concentration in protein and lipid 1 and
a phase with a low concentration in protein and lipid 1.
Using continuation methods
we computed stationary solutions
for model~II. 
Fixing the wavelength to 120\,\textmu{}m (comparable to the
size of a eukaryotic cell) one can also identify
regions, where a linearly stable uniform solution coexists with a linearly stable
stationary periodic solution. 

%
\begin{figure}[hbt]
\includegraphics[width=0.7\hsize]{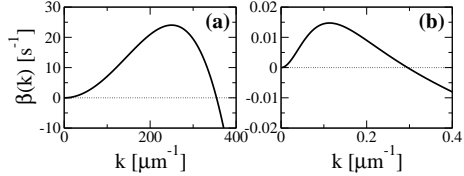}
\caption{Dispersion relations for models~I (a) and II (b) for the uniform
solution. Shown is only the largest (real) eigenvalue. Parameters in (a)
are $N=25$, $c_l=0.2$, $c_m=0.5$, $u=1.01$,
$D_l=1.0\,\mu$m$^2$\,s$^{-1}$ \cite{Kusba:2002}, $D_m=0.04\,\mu$m$^2$\,s$^{-1}$,
$\chi=10^{-6}\,\mu$m$^2$ and in (b) $c_t=0.5$,
$k_{ki}=0.05$\,s$^{-1}$. Remaining parameters are as in Fig.\,\ref{phase}\,(b).\label{disp}} 
\end{figure}

%
%
%
Figs.\,\ref{disp}\,(a) and (b) show the largest eigenvalues for both models from the linearly
unstable regions in Figs.\,\ref{phase}\,(a) and (b).
The instabilities belong
to the type II$_s$ class \cite{Cross:1993}, which is characterized by a real critical
eigenvalue with wave number zero and can be attributed to the conservation relations in
both models. 
Although both models have instabilities of the same type they develop
on very different length and time scales. 
The wavelength of the fastest growing mode $\lambda_m$ in model~I is linked to
the molecular interaction length. 
In our example in Fig.\,\ref{disp}\,a it is
on the scale of 50\,nm with a growth rate of 10\,s$^{-1}$. 
In contrast, $\lambda_m$ of model~II is
determined by kinetic rate and diffusion constants and is of the order of
10\,\textmu{}m. 
In the example in Fig.\,\ref{disp}\,b the corresponding growth
rate is 0.01\,s$^{-1}$.
%
%
\begin{figure}[hbt]
\includegraphics[width=0.7\hsize]{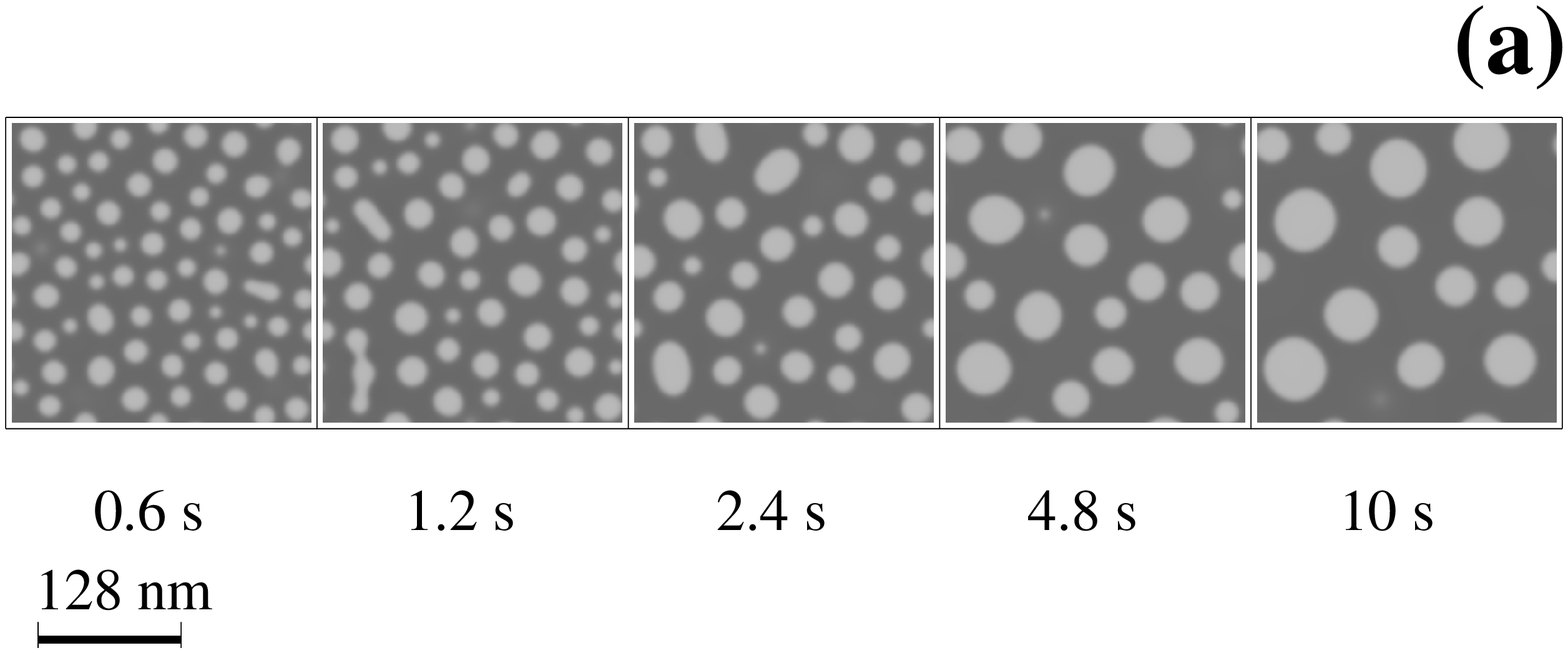}\\
\includegraphics[width=0.7\hsize]{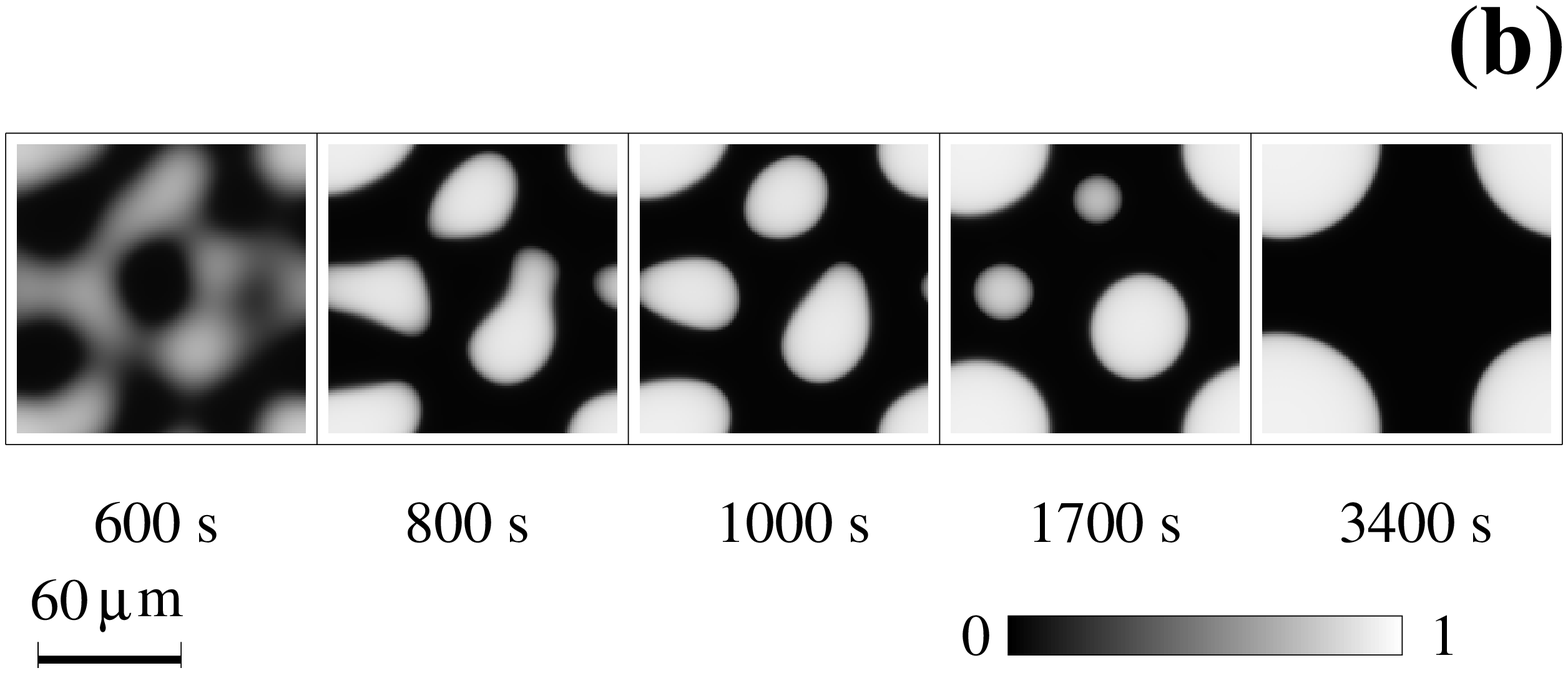}
\caption{Snapshots of a two-dimensional numerical simulations of models~I (a) and II (b) with periodic boundary conditions. Shown is the concentration of membrane
bound protein $c_m$. Parameters in (a) and (b) are as in
Fig.\,\ref{disp}\,(a) and (b), respectively.\label{sim}} 
\end{figure}
The results of the linear stability were confirmed by numerical simulations in
two dimensions with periodic boundary conditions. Simulations were started
from the uniform steady state with small amplitude perturbations.
Figs.\,\ref{sim}\,(a) and \ref{sim}\,(b) show exemplary two simulations for models~I and II with parameter values from the linearly unstable
regions in Fig\,\ref{phase}.
In both models stationary structures develop, which are not stable but display a coarsening
behavior for later stages.
\begin{figure}[hbt]
\includegraphics[width=0.7\hsize]{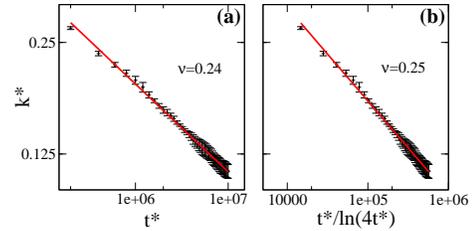}\\[1ex]
\caption{Late stage behavior of model~I ($t^\ast=tD_l/\chi$, $k^\ast=k\sqrt{\chi}$). Shown are loglog-plots of the
evolution of the mean wave number of the structure factor of the membrane
bound protein concentration with time (a) or a modified time (b) as described
in the text. The exponent $\nu$ of the power law was obtained by a least
square fit. Parameters are the same as in
Fig.\,\ref{disp}\,(a).\label{scal}}. 
\end{figure}
For model~I we found the scaling law $\bar{k}(t)=at^{-\nu}$ (Fig.\,\ref{scal})
with an
exponent $\nu\approx{}1/4$ consistent with a modified Lifshitz-Slyosov-Wagner
theory \cite{Bray:1995} for concentration dependent mobility coefficients.
Since we are considering the two-dimensional case one should rather expect a
growth law $\bar{k}(\tau)=a\tau^{-\nu}$ with a modified time scale
$\tau=t/\ln(4t)$ \cite{Rogers:1989}. However, on the time scale of our
numerical experiments both growth laws yielded
similar exponents.
Although the initial growth is very fast the initally observed wave
numbers are small. To obtain structures on the scale of the cell coarsening
has to occur over several orders of magnitude with $\nu=1/4$, which is a slow process. 
The possible scaling behavior of model~II is irrelevant for practical purposes
since the initially developed structures are already on a  scale comparable
to the system size and the first or second coarsening step will lead to
polarized cells with a single domain of high concentration of membrane
bound proteins.
One can easily see in Fig.\,\ref{sim}\,(b) that large structures have appeared after ten minutes and
within one hour coarsening is complete. 

%
%
%
In this Letter we have introduced and analyzed two alternative models for
pattern formation of GMC proteins.
GMC proteins are on the one hand found to form domains by virtue of their
attractive electrostatic interaction with acidic lipids from self-assembly.
On the other hand, they can exploit an ATP-driven
phosphorylation-dephosphorylation cycle (myristoyl-electrostatic switch) for
their self-organization.
The striking difference between both mechanisms lies in the relevant time and
length scales.
The spinodal length scale of the phase separation in model~I is
closely linked to the molecular interaction length in the lipid phase.
For fluid membranes under physiological conditions the relevant interaction scale is 
comparable to the size of a lipid molecule ($\approx 1\,$nm).
The initial structure formation is fast with growth rates of the order of
10\,s$^{-1}$, but the coarsening process follows a scaling law.
Thus coarsening does not lead to structures on the size of the cell in a
biologically relevant time.
The reaction-diffusion mechanism (model~II) leads initially to large
structures, which are on the scale of an eukaryotic cell.
A rough estimate for the length scale is given by the quantity $\sqrt{D \tau}$, 
where $D\approx{}$10\,\textmu{}m$^2$\,s$^{-1}$ is the typical intracellular diffusion
constant of a protein and $\tau =10\,$s is a typical time for a
biochemical reaction.
This yields a length scale of $\approx{}$10\,\textmu{}m.
A combination of both mechanism leads to a more complicated model, that
displays oscillatory dynamics and traveling domains  \cite{John:2005}.
Here, we have used specific physical and chemical properties of
GMC proteins, but the typical scales for molecular interaction energies
and ranges as well as for reaction rates and diffusion constants will be
comparable for other processes near membranes.
We propose that both mechanisms are relevant for different
aspects of structuring membranes:
protein-lipid interactions are suitable for rapid structuring
of membranes on a submicrometer scale, whereas reaction and diffusion of
proteins produce a structure on the scale of the size of typical
eukaryotic cell within minutes and are potentially useful
for polarizing a whole cell into two main compartments (e. g. front and
back).
%
%
%

\begin{thebibliography}{10}
\bibitem{par:u}
The contribution of the protein-lipid interactions to the electrostatic
  energy of the membrane is $-u=\lambda_D
  Q_mQ_l/a_l\epsilon_0\epsilon{}N\approx{}-4\,k_BT$, with $\lambda_D=1$\,nm,
  $Q_m=11\,e$ (charge of proteins), $Q_l=-1\,e$ (charge of type~1 lipids),
  $a_l=1$\,nm$^2$ (area of lipid headgroup), $\epsilon=80$ and $N=25$.
\end{thebibliography}
%
%

%
\end{document}